\pdfoutput=1
\documentclass[twoside,twocolumn,9pt]{extarticle}

\usepackage[T1]{fontenc}
\usepackage{geometry}
\usepackage{graphicx}
\usepackage{amsmath,amssymb}
\usepackage{bm}
\usepackage{enumitem}
\usepackage{subcaption}
\usepackage{float}
\usepackage{dblfloatfix}   

\usepackage[percent]{overpic}
\usepackage[super,sort&compress]{natbib}   
\setcitestyle{citesep={,}}                 
\usepackage{times}
\usepackage[hidelinks]{hyperref}

\begin{document}

\twocolumn[{%
\begin{center}
\vspace*{0.4cm}
{\LARGE\bfseries Valency-bounding correction potential for coarse-grained
molecular dynamics simulations\,$^{\dag}$}\\[2.2ex]
{\large Vladimir Dmitriev$^{\ast ab}$, Ankit Gupta$^{ab}$ and Anton Goloborodko$^{\ast a}$}\\[1.6ex]
\end{center}
\vspace{0.2cm}
\noindent\rule{\textwidth}{0.4pt}\\[0.4ex]
{\noindent\small

Many systems in soft and living matter bind through a limited number of bonds per
particle: proteins associate via discrete surface patches, nucleic acids form
one-to-one contacts, and the phase behaviour of multivalent biomolecules is governed by
the number of binding sites they carry. In simulations, valency limits are typically
enforced with patchy particles, whose anisotropic potentials require precise
integration of rotational degrees of freedom and combine hard cores with narrow patches
to prevent multi-partner binding, which forces small timesteps and restricts
performance and scalability. Patches also commit the model to a fixed binding-site
geometry that is often unknown, flexible or mobile. We introduce the valency-bounding
correction (VBC), a many-body modification of generic short-range pairwise potentials
that smoothly suppresses attraction once the neighbour count of either interacting
particle exceeds a prescribed valency. The correction carries no angular degrees of
freedom, applies on top of soft repulsive cores and evaluates in two passes over the
neighbour list at the cost of a standard pairwise potential. The VBC drives the
coordination number to the prescribed valency with low error, while its cluster
statistics depart from Wertheim and Flory--Stockmayer predictions through unrestricted
ring formation, suggesting qualitatively distinct assembly behaviour for systems with
flexible or surface-mobile binding sites. A tuned variant of the correction exchanges
bonded partners through ordinary molecular dynamics, reducing bond lifetimes at high
saturation by an order of magnitude. On GPUs the cost of the correction is nearly
independent of valency, reaching an almost tenfold advantage over the patchy-particle reference.
We anticipate that the VBC will enable large-scale, top-down-calibrated simulations of
multivalent systems, from associating polymers and gels to biomolecular condensates,
which we illustrate by reproducing the reentrant aggregation of repeat-expanded RNA at
a fraction of the cost of finer-grained models. Additionally, we show how the VBC can
be used to remedy the Fisher--Ruelle thermodynamic instability of soft-core potentials
with attraction. The VBC is available as an open-source GPU plugin for HOOMD-blue.\\[0.4ex]}
\rule{\textwidth}{0.4pt}\\[1.2ex]
}]

\noindent{\footnotesize $^{a}$~Institute of Molecular Biotechnology of the Austrian Academy
of Sciences (IMBA), Vienna BioCenter (VBC), Dr.~Bohr-Gasse 3, 1030 Vienna, Austria.}\par
\noindent{\footnotesize $^{b}$~Vienna BioCenter PhD Program, Doctoral School of the
University of Vienna and Medical University of Vienna, A-1030 Vienna, Austria.}\par
\noindent{\footnotesize\raggedright $^{\ast}$~Corresponding authors. 
E-mail: \mbox{vladimir.dmitriev@imba.oeaw.ac.at}, \mbox{anton.goloborodko@imba.oeaw.ac.at}\par}
\noindent{\footnotesize $\dag$~Electronic supplementary information (ESI) available
as an ancillary file of this preprint.}\par

\section{Introduction}\label{sec:intro}

Coarse-grained molecular dynamics (MD) models have proved useful across soft and
living matter: from colloids~\cite{pusey1986phase,auer2001prediction} and
polymer melts, micelles and
membranes~\cite{kremer1990dynamics,muller2002coarse,marrink2007martini},
to protein conformational changes~\cite{kmiecik2016coarse}, chromatin folding inside
nuclei~\cite{mirny2011fractal,nuebler2018chromatin,goloborodko2016chromosome} and phase separation inside cells in
general~\cite{espinosa2020liquid,brangwynne2015polymer}.

Although the choice and parametrisation of a coarse-grained model remain a
challenge~\cite{saunders2013coarse,ingolfsson2014power}, the qualitative behaviour of such models is
often robust to the functional form of the interaction potentials: for short-ranged
attractions, systems with matched second virial coefficients share the same phase
behaviour and structure~\cite{noro2000extended,vliegenthart2000predicting,platten2015extended}. This justifies the
use of semi-empirical pairwise potentials calibrated top-down against experimental
mesoscale observables~\cite{noid2013perspective,brini2013systematic}.

A particularly widely applicable class of coarse-grained potentials combines a hard or
soft repulsive core with a short-range
attraction~\cite{likos2001effective,zaccarelli2007colloidal}. Its most common members are built on
the Lennard-Jones form~\cite{jones1924determination} or on a separation of the purely repulsive
core from the attractive tail~\cite{weeks1971role}, with the attraction short-ranged compared
to the particle size~\cite{noro2000extended}. They inherit the properties of atomic potentials,
in particular spherical symmetry and the absence of bond
saturation~\cite{likos2001effective,bianchi2011patchy}.

However, many systems of interest in soft and living matter are not well described by
non-saturating spherically symmetric potentials. Globular proteins associate through a
small number of localised binding patches rather than over their entire
surface~\cite{mcmanus2016physics,nguemaha2018liquid}. Nucleic acids form specific one-to-one
Watson--Crick contacts, which coarse-grained models must encode explicitly to reproduce
hybridisation thermodynamics~\cite{ouldridge2011structural,vsulc2012sequence}. The phase behaviour of
multivalent biomolecules is controlled directly by the number of binding sites they
carry~\cite{li2012phase}. These phenomena can be approached with MD techniques allowing for
non-spherical geometry of the interacting
particles~\cite{glotzer2007anisotropy,damasceno2012predictive} or a limited number of bonds each of them can
form~\cite{kern2003fluid,bianchi2011patchy}. The latter is typically achieved through
patchy-particle models~\cite{kern2003fluid,bianchi2011patchy}.

The patchy-particle approach has been successfully used to study phase transitions in
limited-valency fluids~\cite{bianchi2006phase,bianchi2008theoretical,romano2010phase,foffi2007possibility}, with
direct applications to soft and living matter
systems~\cite{nguemaha2018liquid,espinosa2020liquid}. Its theoretical foundation rests on
Wertheim's theory of highly directional associating
fluids~\cite{wertheim_1_1984fluids,wertheim_2_1984fluids,wertheim_1_1986fluids, wertheim_2_1986fluids},
developed further through the statistical associating fluid theory
(SAFT)~\cite{chapman1989saft}.

Despite their wide applicability and theoretical foundation, patchy particles carry
limitations that become critical in large-scale studies, in particular in the extensive
parameter scans that top-down calibration requires.

The first limitation is computational. Patchy-particle models run in Monte Carlo (MC) or MD
frameworks, and MC lacks parallel scalability, while access to modern GPUs is what
makes the equilibration of large systems, such as polymer solutions or gels,
affordable. Parallel MC algorithms have been developed~\cite{anderson2013massively}, but they
mostly do not extend to polymeric or gel-forming systems. MD parallelises well and
scales across GPU clusters~\cite{anderson2008general,glaser2015strong,anderson2020hoomd}. 
Patchy-particle MD implementations, however,
require a hard repulsive core and patches of restricted size. The resulting steepness of
the potential degrades numerical stability and forces a small integration
timestep~\cite{rovigatti2018simulate}. Beyond numerics, patchy MD integrates rotational degrees
of freedom and therefore resolves the orientational dynamics of the patches. When this
dynamics is irrelevant to the observables of interest, computational resources are spent
on a timescale the study does not need.

Another, subtler issue is that the behaviour of a patchy system is not robust with
respect to how the patches are arranged on the particle
core~\cite{zhang2004self,bianchi2008theoretical,chen2011directed,wang2012colloids}. The dependency of
physical properties on patch positions may be beneficial if the positions of the
interaction sites on a real molecule are known. However, if this information is not
available, an arbitrary choice of patch positions becomes an uncontrolled model assumption.
 Moreover, a fixed arrangement may misrepresent the physics when the
binding sites are flexible polymer chains~\cite{arya2006role}, or when binding is
mediated by a third component that is mobile on the particle
surface~\cite{van2013solid,angioletti2014mobile}.

To address these limitations, we propose a valency-bounding correction (VBC) to pairwise
interaction potentials that limits the effective coordination number of particles.
The VBC extends the family of many-body interatomic potentials.
For each particle it aggregates the local neighbour
count into a scalar per-particle sum, inheriting this idea from the embedded-atom
method~\cite{daw1984embedded}. The VBC suppresses the attractive component of the potential 
when the smoothed neighbour count exceeds a prescribed threshold. 
A similar construction of the smoothed neighbour count
reappears in coarse-grained many-body interaction models, for instance to achieve
liquid--vapour coexistence with dissipative particle
dynamics~\cite{pagonabarraga2001dissipative,warren2003vapor}. 

The VBC imposes no constraint on the positions of bonds, only on their number. Unlike
bond-order and three-body forms of multi-body
interactions~\cite{stillinger1985computer,tersoff1988new,brenner2002second}, our correction introduces no
angular dependence and needs no explicit evaluation over all triplets in the system. That
guarantees computational efficiency --- the correction for a pair depends only on the
distances of the other neighbours to the two particles in the pair. Force evaluation
scales as $O(kN)$ with the neighbour-list size~$k$, matching the cost of standard
pairwise potentials.

In Section~\ref{sec:potential} we formally introduce the valency-bounding correction
(VBC) and its implementation. In Section~\ref{sec:binding} we show the binding behaviour
and the resulting cluster-size distributions, outlining differences from classical
patchy-particle behaviour. In Section~\ref{sec:lifetimes} we study the dynamics of bonds
and show that the VBC and patchy particles demonstrate comparable rates of neighbour
exchange. Moreover, we show adjustments of the VBC facilitating faster swapping of
neighbours. In Section~\ref{sec:performance} we benchmark the computational cost of the
correction against patchy particles, demonstrating the high efficiency of the VBC.
Finally, in Section~\ref{sec:applications} we present two applications of the correction:
reproducing a reentrant phase transition in RNA mixtures~\cite{kimchi2023uncovering}, and restoring
the thermodynamic stability of soft-core potentials with attractive terms.

\section{VBC definition}\label{sec:potential}

This section defines the correction for a generic pairwise potential and then
specialises it to the polynomial soft-core potential used throughout the paper.

\subsection{General formulation}\label{subsec:general}

\begin{figure}[t]
    \centering
    \includegraphics[width=\columnwidth]{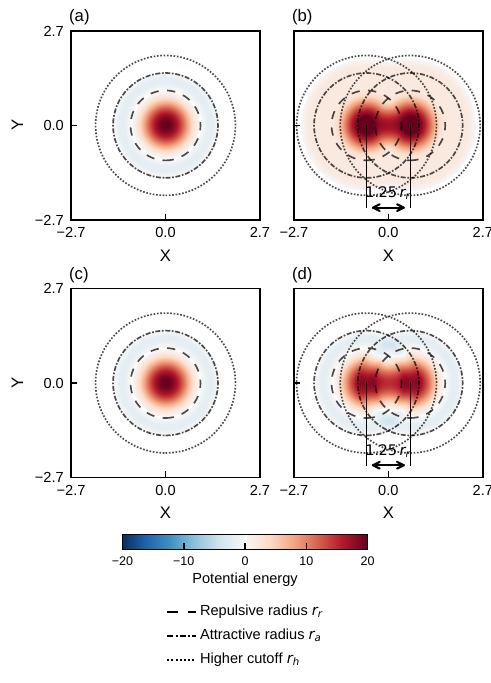}
    \caption{\label{fig:val_heatmaps}
    Potential energy landscape with valency control for the standard VBC with the
    polynomial soft-core potential ($r_r=1$, $r_a=1.5$, $E_a=2$, $E_r=20$).
    Colour is the insertion energy of the probe particle, i.e.\ the change in the total
    potential energy $\sum_{i<j}\hat{V}(r_{ij})$ upon adding the probe.  Red denotes
    repulsion, blue attraction.
    Each row fixes the target valency: (a,\,b)~$f=1$, (c,\,d)~$f=2$.
    Each column fixes a configuration: (left)~a single fixed neighbour and (right)~two
    fixed neighbours separated by $1.25\,r_r$.  Dashed, dash-dotted and dotted circles
    mark the repulsive radius~$r_r$, the attractive radius~$r_a$ and the higher
    cutoff~$r_h$.  When the neighbour count reaches the target valency, the attractive
    well is suppressed for the probe particle.}
\end{figure}

We build the VBC on a general pairwise potential that decomposes into a repulsive core and an attractive shell:
\begin{equation}
V(r_{ij}) = V_r(r_{ij}) + V_a(r_{ij}),
\end{equation}
where $V_r \ge 0$ is the repulsive part with support in
$(0,\, r_r)$ and $V_a \le 0$ is the attractive part
with support in $(r_r,\, r_a)$.

The VBC is based on a two-step calculation: the first step evaluates the smoothed
neighbour count around each particle, and the second multiplies the attractive part of
the potential by a correction factor that depends on the smoothed neighbour counts of the
two interacting particles.
For both steps we use a smooth step function, defined as
\begin{equation}
\omega(x, x_1, x_2) =
\begin{cases}
1, & x \le x_1,\\[4pt]
\biggl(1 - \Bigl(\dfrac{x - x_1}{x_2 - x_1}\Bigr)^{\!2}\,\biggr)^{\!2}, & x_1 < x < x_2,\\[10pt]
0, & x \ge x_2.
\end{cases}
\label{eq:omega}
\end{equation}
This function is continuously differentiable, ensuring continuous forces.
Using it, we define the smoothed neighbour count of particle~$i$ as
\begin{equation}
n_i = \sum_{j \neq i}^{N}
  \omega\bigl(r_{ij},\, r_\ell,\, r_h\bigr).
\label{eq:counter}
\end{equation}
A neighbour contributes~1 while $r_{ij} < r_\ell$, and the contribution decays to zero
between the lower cutoff $r_\ell$ and the higher cutoff~$r_h > r_\ell$. Unless stated
otherwise, we choose $r_\ell = r_a$, which guarantees that the count contribution of each
neighbour is exactly~1 as soon as the neighbour enters the attraction range. Choosing
$r_h$ well above $r_\ell$ avoids steep gradients in the force. At the same time, setting
$r_h$ too large creates effective long-range interactions and forces a larger
neighbour-list cutoff.

The corrected pairwise potential is then
\begin{equation}
\hat{V}(r_{ij}) = V_r(r_{ij})
+ \Psi\bigl(n_i,\, n_j\bigr)\, V_a(r_{ij}),
\label{eq:Vhat}
\end{equation}
where the switching function
\begin{align}
\Psi(n_i, n_j) &= \omega\bigl(n_i,\, f,\, f{+}1\bigr)
\notag \\
&\quad{}\times
\omega\bigl(n_j,\, f,\, f{+}1\bigr)
\label{eq:Psi}
\end{align}
uses $\omega(x, x_1, x_2)$ to suppress the attractive component as soon as the
neighbour count around either particle exceeds $f$, nullifying it completely once the
count reaches $f+1$.
The constant $f$ specifies the target upper limit on the number of neighbours
around a particle.

Both the energy and the force (Section~S1 of the ESI\,$\dag$) depend on the environment
of $i$ and $j$ only through the scalar counts $n_i$ and $n_j$. Two passes over the
neighbour list therefore suffice: the first accumulates the counts, the second evaluates
energies and forces, including the chain-rule contributions through
$\partial\Psi/\partial n$. Evaluation then scales as $O(kN)$ with the neighbour-list
size~$k$, matching a standard pairwise potential and avoiding the $O(k^2N)$ cost of an
explicit loop over triplets. We implemented the VBC as a GPU-accelerated
HOOMD-blue~\cite{anderson2020hoomd} plugin.

The VBC generalises to multi-component systems, in which each type pair
carries its own valency cap and interaction parameters, the counting radii are
symmetric in the pair, and the count of a given particle can be restricted to
selected neighbour types. The general multi-type formulation is given
in Section~S1 of the ESI\,$\dag$.

Throughout the paper, the valency-bounding correction is applied to the polynomial
soft-core potential developed for coarse-grained chromatin
simulations~\cite{samejima2025rules,corsi2025conformational}. 
The potential is
\begin{align}
V_r(r_{ij}) &= E_r
\Bigl(1 - \bigl(r_{ij}/r_r\bigr)^2\Bigr)^{2}\,
\theta\bigl(r_r - r_{ij}\bigr),
\notag\\[6pt]
V_a(r_{ij}) &= -E_a
\Bigl(1 -
  \Bigl(
    \frac{r_{ij} - \frac{1}{2}(r_a+r_r)}
          {\frac{1}{2}(r_a-r_r)}
  \Bigr)^{\!2}
\Bigr)^{2}
\notag\\
&\quad\times
\theta\bigl(r_{ij} - r_r\bigr)\,
\theta\bigl(r_a - r_{ij}\bigr),
\label{eq:softcore}
\end{align}
where $E_r > 0$ and $E_a > 0$ set the height of the repulsive barrier and the depth of
the attractive well, $r_r$, $r_a$ set the corresponding distance scales, and $\theta$ is
the Heaviside step function.  The quartic form makes the potential $C^1$-continuous,
ensuring continuous forces, and the soft core permits large time steps.

We refer to the correction with $r_\ell = r_a = 1.5\,r_r$ and $r_h = 2\,r_r$ applied to
this polynomial potential as the standard VBC. It is used throughout the paper
unless stated otherwise.  Figure~\ref{fig:val_heatmaps} illustrates the effect of the
standard VBC for target valencies~$f=1$ and~$f=2$.

Throughout the paper, unless stated otherwise, all simulations use $N=10\,000$ particles. 
For the VBC we use the polynomial potential
with $r_r=1$, $r_a=1.5$, $r_\ell=r_a$, $r_h=2\,r_r$, $E_r=20$, $E_a=2$, time step
$dt=0.01$, and Langevin damping $\gamma=0.1$ in reduced units.  Each run covers $10^7$ integration steps.  
Full simulation parameters are given in Section~S3 of the ESI\,$\dag$.

\section{Binding and cluster distributions}\label{sec:binding}

\begin{figure}[t]
    \centering
    \includegraphics[width=\columnwidth]{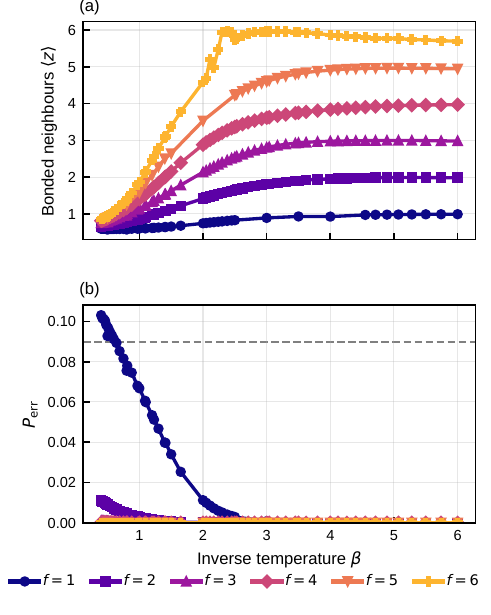}
    \caption{\label{fig:saturation}
    Coordination and its error versus inverse temperature $\beta$ for prescribed
    valencies $f=1$--$6$, for the standard VBC at $E_a=2$ and $\rho=0.05$.
    (a)~The average number of bonded neighbours $\langle z\rangle$ rises sigmoidally as
    the fluid is cooled and saturates on a plateau at its prescribed valency, $\langle
    z\rangle\to f$.  (b)~The over-coordination probability $P_\mathrm{err}$: the fraction
    of particles with more than $f$ neighbours. The dashed line marks the expected level of random transient passages through the attractive shell.}
\end{figure}

We begin by testing whether the VBC does what it was developed for, namely holding the coordination number of particles at the
prescribed valency. We simulate fluids of
particles with fixed valencies $f=1$--$6$ across a range of temperatures and follow two
observables. The first is the average number of bonded neighbours per particle,
$\langle z\rangle$. The second is the fraction of particles carrying more than $f$
neighbours, $P_\mathrm{err}$. Throughout, two particles count as neighbours when their
separation lies within the attraction range, $r_{ij} < r_a$ --- an instantaneous
distance criterion, so transient contacts are counted alongside bonds.

Figure~\ref{fig:saturation}a follows $\langle z\rangle$ as the fluid is cooled. As the
inverse temperature $\beta$ grows, the binding strength in thermal units, $\beta E_a$,
increases and $\langle z\rangle$ rises steeply and sigmoidally. Each curve approaches a
plateau at its prescribed valency, $\langle z\rangle \to f$, for every $f$.

$P_\mathrm{err}$ stays small across the whole temperature range
(Fig.~\ref{fig:saturation}b). It is largest for $f=1$ at weak binding and decays to zero
on cooling. The residual excess in this regime comes from random transient passages of particles
through the attractive shell, whose expected level is marked by the dashed line in
Fig.~\ref{fig:saturation}b. At $\rho=0.05$ a simple estimate gives on
average ${\approx}0.5$ shell occupants per particle and a ${\approx}9\%$ chance of two
or more simultaneous occupants (Section~S6 of the ESI\,$\dag$), which sets the scale of
the measured $P_\mathrm{err}$ for $f=1$ at high temperature. Since the penalty for
over-coordination grows with both the binding strength $\beta E_a$ and the number of
neighbours $z$, the correction at $f=1$ and high temperature is too weak to deflect
thermal collisions. On cooling the penalty steepens, random encounters are suppressed,
and $P_\mathrm{err}$ decays to zero.

The VBC thus enforces the prescribed valency with an error that is bounded by random
shell traffic at weak binding and vanishes in the strongly bound regime, where valency
control actually matters.

\subsection{Saturation and connectivity}\label{subsec:saturation}

Next we study how the saturation $p=\langle z\rangle/f$, the occupied fraction of a particle's bond
slots, changes with valency and temperature. 
In the theory of associating
fluids~\cite{wertheim_1_1984fluids, wertheim_2_1984fluids, wertheim_1_1986fluids, wertheim_2_1986fluids}, the saturation
of a patchy particle equals the probability of a patch forming a bond, provided that
binding events are independent.

Alongside the saturation we track connectivity through the fraction of particles in the
giant component, $P_\infty$, which we use as a practical proxy for the onset of phase
separation. We do not determine the phase boundaries directly, and $P_\infty$ probes the
mass of the largest cluster rather than the presence of a system-spanning network. At
densities well below the critical one, however, the percolation locus and the binodal of
short-range attractive fluids lie close to each other, and a macroscopic cluster emerges
through condensation rather than through connectivity of the homogeneous
fluid~\cite{miller2003competition, lu2008gelation}. Consistently, at low temperatures we directly observe
dense droplets (Section~\ref{sec:clusters}).

\begin{figure}[t]
    \centering
    \includegraphics[width=0.95\columnwidth]{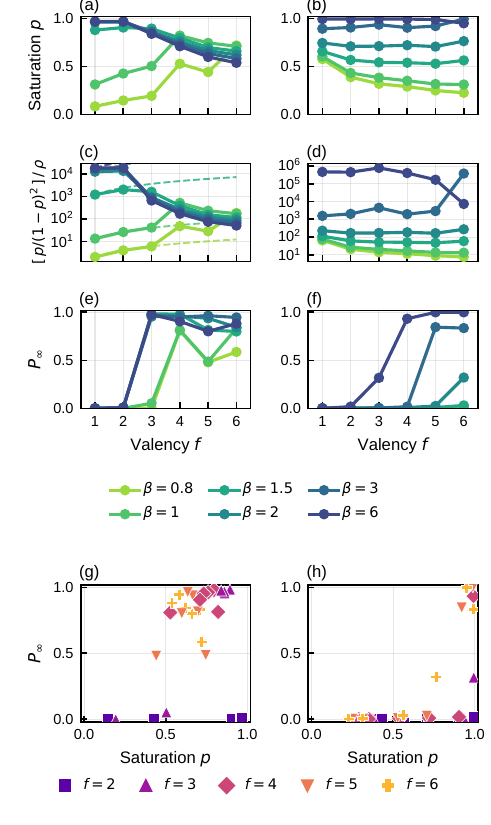}
    \caption{\label{fig:massaction}
    Saturation, the law of mass action, and connectivity, at $\rho=0.05$, for the
    patchy-particle reference ($E_a=12$, Section~S5 of the ESI\,$\dag$, left column) and
    the standard VBC ($E_a=2$, right column).
    (a,\,b)~Saturation $p$.
    (c,\,d)~The law-of-mass-action combination $[\,p/(1-p)^2\,]/\rho$ of
    Eq.~\eqref{eq:massaction} against valency on log-$y$ axes.
    (e,\,f)~The giant component fraction $P_\infty$ against valency, marking
    the onset of aggregation, used here as a proxy for phase separation.
    (g,\,h)~The same $P_\infty$ against the saturation $p$.
    In panels (a)--(f) colour denotes inverse temperature $\beta$, valencies span
    $f=1$--$6$. In panels (g,\,h) each point is a per-$(f,\beta)$ median
    coloured by valency $f=2$--$6$.}
\end{figure}

For the patchy-particle reference in the sol regime, where clusters
remain finite and predominantly tree-like, the law of mass action connects the
saturation $p$ with the density $\rho$ and the valency $f$ for a given shape of the
attraction well~\cite{wertheim_1_1984fluids, wertheim_2_1984fluids, wertheim_1_1986fluids,wertheim_2_1986fluids}:
\begin{equation}
\frac{p}{(1-p)^{2}} = f\,\rho\,\Delta,
\qquad
\Delta = \int \bigl[\,e^{-\beta V_b(\mathbf r)} - 1\,\bigr]\,
g_{\mathrm{ref}}(r)\,\mathrm{d}\mathbf r,
\label{eq:massaction}
\end{equation}
where $\Delta$ is the integral of the Mayer function of a single attraction site $V_b$
over the pair correlation function $g_{\mathrm{ref}}$ of the reference fluid. At fixed
temperature and interaction parameters the combination
$[\,p/(1-p)^{2}\,]/\rho = f\,\Delta$ thus grows linearly with the valency, provided that
all patches are equivalent and the patch size does not depend on their number. The
patchy fluid follows this prediction closely. The saturation grows with valency
(Fig.~\ref{fig:massaction}a) and the mass-action combination rises along the expected
linear form (Fig.~\ref{fig:massaction}c). Deviations appear at low temperatures and
coincide with the formation of a giant component (Fig.~\ref{fig:massaction}e). The same
onset shows up as a jump in the saturation curves of Fig.~\ref{fig:massaction}a.

The VBC fluid departs from this picture --- its saturation is almost flat in valency
(Fig.~\ref{fig:massaction}b), and the mass-action combination stays nearly constant
instead of growing linearly (Fig.~\ref{fig:massaction}d). These deviations are
independent of the formation of a giant component. Two properties of the correction
explain the difference. First, all bonds of a VBC particle share the same binding
volume. Increasing the valency at fixed $E_a$ adds bond slots without adding attraction
volume, so additional neighbours compete for the same shell and the saturation per slot
stays flat. Reproducing the patchy-like growth of saturation with valency therefore
requires an adjustment of the binding energy alongside the valency. 
An empirical occupancy model that reproduces the measured saturation across valencies
 is given in Section~S2 of the ESI\,$\dag$.

The giant component for the VBC forms only at larger valencies and lower temperatures
than in the patchy reference (Fig.~\ref{fig:massaction}f). This late formation, however,
cannot be explained by the binding-volume competition. As panels (g,\,h) show, the VBC
giant component stays negligible for most valencies up to high saturation, whereas the
patchy-particle reference aggregates substantially already from $p\approx0.5$.
This difference cannot be attributed to bonded neighbours sharing the same binding volume, 
because the comparison is made at equal $p$. 
High saturation in the absence of a
giant component implies that particles form most of their bonds within small clusters.
As we show, this is the second property of the correction: VBC particles readily form
cyclic clusters, which are far rarer for patchy particles and often absent from their 
analytical treatment.

\subsection{Cyclisation}\label{subsec:cyclisation}

\begin{figure}[t]
    \centering
    \includegraphics[width=0.95\columnwidth]{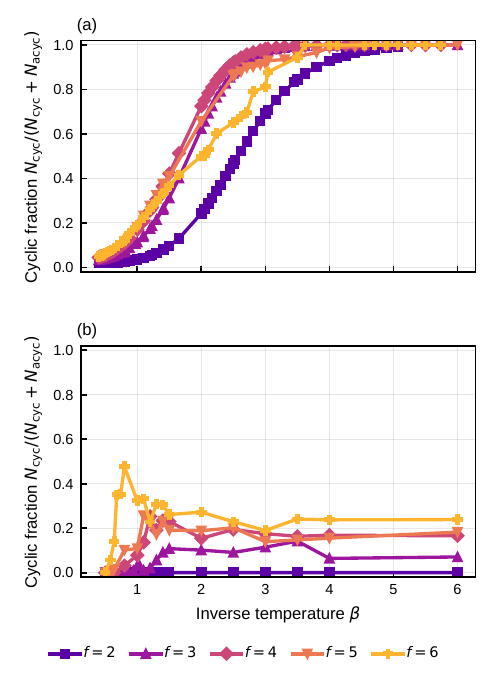}
    \caption{\label{fig:cyclic}
    Fraction of particles in cyclic clusters,
    $x_\mathrm{cyc}=N_\mathrm{cyc}/(N_\mathrm{cyc}+N_\mathrm{acyc})$, versus inverse
    temperature $\beta$ at $\rho=0.05$, for (a)~the standard VBC ($E_a=2$) and (b)~the
    patchy reference ($E_a=12$, Section~S5 of the ESI\,$\dag$).  Colour denotes valency
    $f=2$--$6$.  The VBC cyclic
    fraction rises smoothly toward unity, whereas the patchy reference saturates far
    below it, reflecting the angular constraints that suppress ring closure.}
\end{figure}

We address the late formation of a giant component for the VBC 
by quantifying the fraction of particles residing in loop-carrying clusters as the temperature is lowered. Figure~\ref{fig:cyclic} tracks the cyclic
fraction $x_\mathrm{cyc}=N_\mathrm{cyc}/(N_\mathrm{cyc}+N_\mathrm{acyc})$, where
$N_\mathrm{cyc}$ and $N_\mathrm{acyc}$ are the numbers of particles residing in
loop-carrying and tree-like clusters, against inverse temperature $\beta$ at
$\rho=0.05$. In the VBC fluid $x_\mathrm{cyc}$ rises smoothly toward unity as bonds
strengthen. In the patchy-particle reference it saturates far below unity --- the
angular constraints suppress ring closure even in a dense network.

\begin{figure*}[b]
    \centering
    \includegraphics[width=0.47\linewidth]{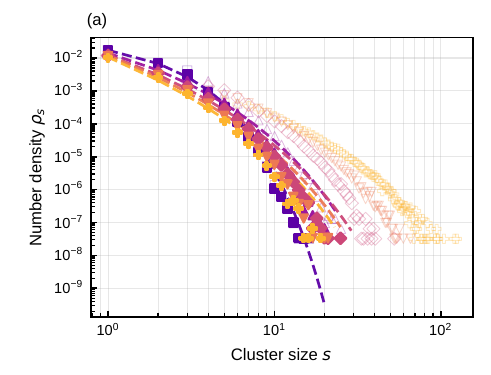}
    \hfill
    \includegraphics[width=0.47\linewidth]{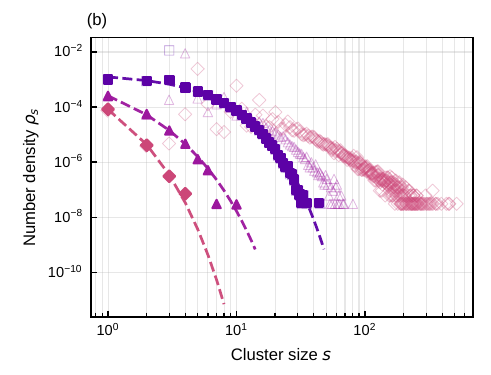}\\[0.4em]
    \includegraphics[width=0.94\linewidth]{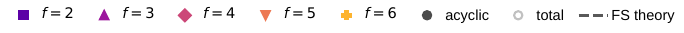}
    \caption{\label{fig:cluster_distributions}
    Cluster-size distributions for the standard VBC ($E_a=2$) at $\rho=0.05$.  Number
    density $\rho_s$ of clusters of size $s$ (log--log).  Filled symbols are acyclic
    clusters, open symbols all clusters including loops, and dashed lines the
    Flory--Stockmayer prediction~\eqref{eq:fs} (evaluated on the acyclic subset).
    Colour denotes valency $f$.
    (a)~$\beta=1$, $f=2$--$6$: the acyclic distributions track FS closely, while the
    total distributions carry an excess at large sizes from ring formation.
    (b)~$\beta=3$, $f=2$--$4$: at this lower temperature the higher valencies $f=5,6$
    have already condensed into dense droplets and are omitted.}
\end{figure*}

The abundance of cyclic clusters explains the observed delay in the formation of the
giant component. When cycles within clusters are allowed, all the binding capacity of a
VBC particle can be saturated within small clusters, whereas for patchy particles ring
closure is suppressed and the formation of a giant component remains the only way to
close cycles and saturate additional bonds. This closes the argument of panels
(g,\,h) of Fig.~\ref{fig:massaction}: the giant-component gap that survives there at
equal saturation is carried by the cyclic clusters.

\subsection{Cluster-size distributions}\label{sec:clusters}

We next investigate how the abundance of cyclic clusters affects the shape of the
cluster-size distribution. The cluster-size distribution of the patchy-particle reference follows
 the classical Flory--Stockmayer (FS) theory of
polymerisation~\cite{flory1941molecular,stockmayer1943theory,flory1953principles} over a wide range of
conditions. For a valency-$f$ fluid with saturation $p$, the FS number density of tree
clusters of size~$s$ is
\begin{equation}
\rho_{s} = \rho\,w_{s}\,(1 - p)^{f}\,
\bigl[p\,(1 - p)^{f-2}\bigr]^{s-1},
\label{eq:fs}
\end{equation}
where
\begin{equation}
w_{s} = \frac{f\,(fs - s)!}{(fs - 2s + 2)!\;s!}.
\end{equation}

With cyclic clusters abundant, the VBC distribution cannot be expected to follow
Eq.~\eqref{eq:fs}. To separate the effect of cycles from the effect of
binding-volume competition, we analyse the acyclic clusters and the full set of clusters
independently. The construction of the acyclic distribution and its normalisation are
detailed in Section~S7 of the ESI\,$\dag$.

Figure~\ref{fig:cluster_distributions} shows both distributions at $\rho=0.05$ for two
inverse temperatures. The acyclic subset largely recovers the FS form, while the full
distribution at large sizes is consistent with the contribution of cyclic clusters. On
cooling to $\beta=3$ the higher valencies $f=5,6$ have already condensed into dense
droplets, so only $f=2$--$4$ are shown.

To summarise, differences between the VBC and patchy particles originate from
the former accommodating all neighbours within the same attractive shell, which 
reduces the binding volume for each next binding partner and allows for easy cycle
formation within a cluster. The binding-volume constraint is of the same nature as for
isotropic potentials and can be corrected for by an adjustment of the attraction energy
within a top-down calibration approach. These properties of the VBC make it particularly fit for modelling systems
 in which limited valency does not arise from short-range 
directional bonds, for instance particles with flexible binding sites or mobile surface-bound 
linkers~\cite{arya2006role,van2013solid,angioletti2014mobile}.

\section{Bond saturation dynamics and lifetimes}\label{sec:lifetimes}

After studying the saturation and clustering of VBC particles at equilibrium, we investigate how fast this
equilibrium is reached. We address two
questions: how fast bonds saturate, and what the characteristic bond lifetime of a VBC
fluid is.

\subsection{Saturation dynamics}\label{subsec:satdynamics}

We study the saturation time of bonds in a fluid of limited-valency particles, comparing
it between the VBC and the patchy-particle reference. To sharpen the comparison in a
nontrivial search scenario, we consider a minimal two-component (A--B) system in which
only the heterotypic attraction is present and the valency cap applies to the same
heterotypic bonds. The composition and simulation protocol are given in Section~S8 of the
ESI\,$\dag$.

\begin{figure}[t]
    \centering
    \includegraphics[width=0.95\columnwidth]{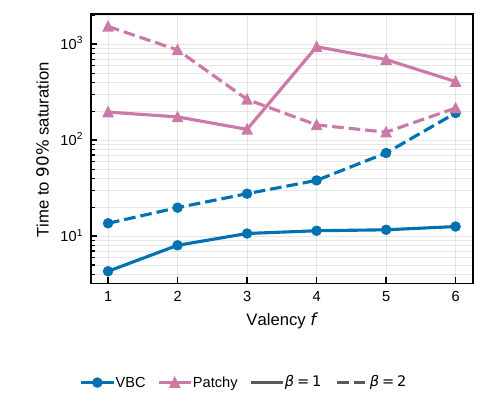}
    \caption{\label{fig:sattime}
    Simulation time for the A--B bond saturation to reach $90\%$ of its plateau value
    (attained after $10^6$ timesteps) versus prescribed valency $f$, at $\rho=0.05$,
    for the standard VBC ($E_a=2$) and the patchy reference ($E_a=12$, Section~S5 of the
    ESI\,$\dag$), at two temperatures ($\beta=1$ solid, $\beta=2$ dashed).  The patchy
    reference saturates close to an order of magnitude more slowly than the VBC.}
\end{figure}

Figure~\ref{fig:sattime} reports this time for the standard VBC and the patchy-particle
reference across valencies and two inverse temperatures. 
Compared to the VBC, the patchy-particle reference needs almost an order of magnitude
longer simulation time. This difference can be attributed to an additional orientational
search patchy particles have to perform to form a patch--patch bond, whereas a VBC bond
forms on contact regardless of orientation.

\begin{figure*}[b]
    \centering
    \includegraphics[width=0.96\linewidth]{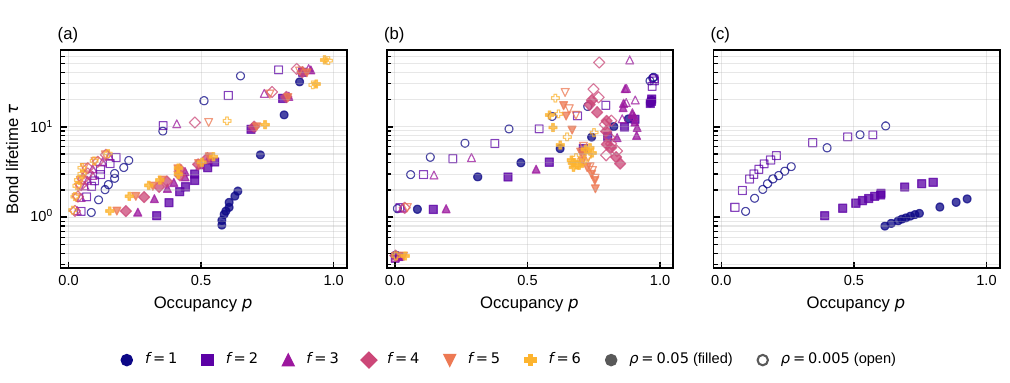}
    \caption{\label{fig:lifetimes}
    Bond lifetime $\tau$ versus saturation $p$ at binding energy $E_a=2$, for (a)~the
    standard $\mathrm{VBC}(1.5,2.0,0)$, (b)~the patchy-particle reference ($E_a=12$,
    Section~S5 of the ESI\,$\dag$) and (c)~the
    optimised $\mathrm{VBC}(1.29,1.50,0.30)$.  Filled markers are $\rho=0.05$ and open
    markers $\rho=0.005$.  $\tau$ is the median bond lifetime over the restart
    trajectory, in units of the timestep~$dt$.  Bond lifetime grows with saturation for
    every model, and the exchange-tuned VBC re-partners fastest at high~$p$.}
\end{figure*}

\subsection{Bond lifetimes}\label{subsec:lifetimes}

Next we address the bond dynamics after saturation is reached. Studies of equilibrium in
fluids or polymers require sampling over times that greatly exceed the individual bond
lifetimes, so that bonds repeatedly break and re-form and the system explores its phase
space. This is crucial for polymeric or gel-forming systems, whose free-energy
landscape with many local minima must be sampled for adequate averaging of observables
and for the discovery of new
conformations~\cite{debenedetti2001supercooled,zaccarelli2007colloidal}. Reaching
equilibrium in such systems has motivated dedicated simulation techniques, most notably
the bond-swap dynamics that lets particles trade partners without arresting the
network~\cite{sciortino2017three}. Since efficient equilibration of large polymeric
systems with limited valencies was the main motivation behind the development of the VBC,
we assess how long individual bonds survive under the correction, and then show that the
correction geometry can itself be tuned to shorten these lifetimes and facilitate faster
bond equilibration.

As a function of saturation, the bond lifetime of the standard VBC (Fig.~\ref{fig:lifetimes}a)
and of the patchy-particle reference (Fig.~\ref{fig:lifetimes}b) rises on a comparable
track as the saturation increases. To reduce bond lifetimes at fixed saturation, we consider optimising the VBC to enhance
the bonded-partner exchange channel. The mechanism of the exchange is built into the VBC,
since the correction weakens attraction as soon as the neighbour count would exceed the
prescribed valency, so the approach of an extra particle to a saturated centre
transiently loosens all of its bonds and lets one partner leave as the new particle takes
its place. A bond is thus handed from one partner to another in a continuous move of the
ordinary MD dynamics. However, in the standard VBC the particle's neighbour count senses 
the newcomer before the attraction does: with $r_h > r_a$, an approaching particle raises the counts, and
thereby the energy, while it is still outside the attractive well. The exchange therefore
proceeds over an energy barrier, which depends on the VBC parameters $r_\ell$ and $r_h$
that we have kept fixed until now. To gain optimisation flexibility, we further generalise
the correction with a shift parameter~$\delta$: the switching function of
Eq.~\eqref{eq:Psi} is evaluated at a threshold displaced from the prescribed valency,

\begin{equation}
\Psi = \omega\bigl(n_i,\, f{-}\delta,\, f{-}\delta{+}1\bigr)\,
       \omega\bigl(n_j,\, f{-}\delta,\, f{-}\delta{+}1\bigr),
\label{eq:Psidelta}
\end{equation}
so that a negative $\delta$ lets the suppression set in slightly after the prescribed
valency is reached, and a positive $\delta$ slightly before. We denote by
$\mathrm{VBC}(r_\ell,\, r_h,\, \delta)$ the correction with counting radii $r_\ell$ and
$r_h$, given in units of $r_r$, and shift~$\delta$. The standard VBC of the previous
sections is $\mathrm{VBC}(1.5,\,2.0,\,0)$ in this notation. The extension gives three
tunable parameters $(r_\ell,\,r_h,\,\delta)$, which reshape the energy profile along
the exchange path and, with it, the barrier separating the bonded states.

For the optimisation we consider a move in which a saturated particle exchanges one
neighbour for another free particle, with all three particles moving along the same line
in space. This path gives a proxy for the exchange barrier height and is sufficient to
compare different VBC geometries. We perform a grid search over the correction parameters
$(r_\ell,\,r_h,\,\delta)$, computing the barrier of the exchange path at every valency
and penalising, alongside the worst-case barrier, any over-coordination trap, a local
minimum of the potential in which an extra particle binds to an already-saturated centre
(Section~S4 of the ESI\,$\dag$). This search gives the optimised
$\mathrm{VBC}(1.29,1.50,0.30)$ (Fig.~\ref{fig:lifetimes}c), for which the exchange
barrier is small and the trap is absent. The coordination curves of this geometry are
given in Section~S4 of the ESI\,$\dag$, and Section~S6 shows that its over-coordination 
probability remains small.

For valencies $f=1$ and $2$ one can use $\mathrm{VBC}(1.29,1.50,0.30)$ with the same
timestep $dt=0.01$ as the standard VBC within the studied temperature range, but for
$f\ge3$ the timestep must be reduced to $dt=0.005$, to avoid integration
errors. We therefore show the optimised $\mathrm{VBC}(1.29,1.50,0.30)$ at $f=1$ and $2$
in Fig.~\ref{fig:lifetimes}c. It keeps bond lifetimes short even at high saturation, in
contrast to the patchy particles, whose lifetimes grow substantially at high $p$.

The optimised VBC thus maintains moderate bond lifetimes even at high saturation $p$,
reaching an order-of-magnitude improvement over both the patchy reference and the
standard VBC at the highest saturations. The exchange channel is provided by the
correction geometry itself, within plain MD, offering an efficient way to equilibrate
strongly bonded limited-valency systems, from reversible gels and
networks~\cite{sciortino2011reversible,zaccarelli2007colloidal} to multivalent biomolecular
mixtures~\cite{espinosa2020liquid,kimchi2023uncovering}.

\section{Computational performance}\label{sec:performance}

Having established that the VBC enforces the prescribed coordination and equilibrates 
faster than the patchy-particle reference, we now analyse its computational cost. We compare 
the VBC and the patchy-particle reference
with the same integrator, neighbour-list backend, system size and state points on
identical hardware, so the difference in wall time reflects the valency-limiting
machinery alone.

\subsection{Cost per integration step}\label{subsec:timing}

We benchmark cost as the wall time needed to integrate $10^6$ MD steps on a single GPU
(NVIDIA A100), with a neighbour-list buffer of $0.4$. The VBC evaluates a scalar
per-particle count at $O(kN)$ cost and carries no rotational degrees of freedom,
whereas the patchy reference integrates rigid-body orientations and evaluates an
angular envelope for every patch pair, so we expect the VBC to be cheaper per step.
The comparison is nevertheless conservative for the VBC, since the standard geometry
requires a neighbour-list radius of $r_h=2\,r_r$, larger than the $1.5\,r_r$ sufficient
for our patchy-particle implementation, and therefore a longer neighbour list.

\begin{figure*}[tb]
\centering
\includegraphics[width=0.96\linewidth]{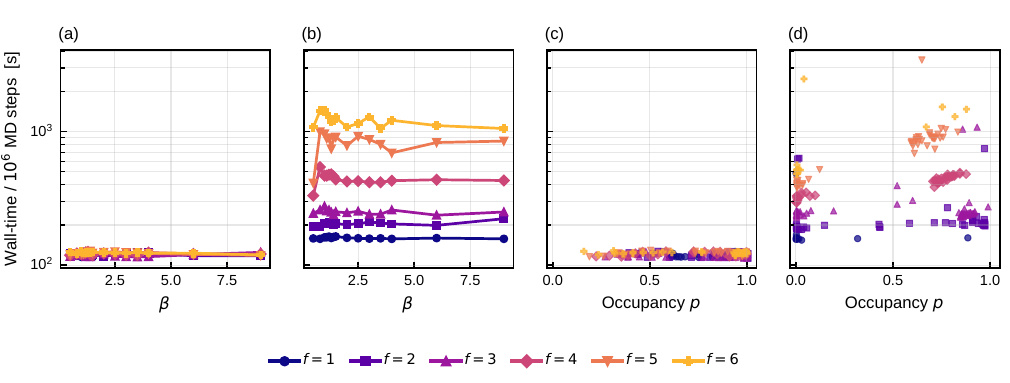}
\caption{\label{fig:wt_cost}
Wall time per $10^6$ MD steps at $\rho=0.05$ for the standard $\mathrm{VBC}(1.5,2.0,0)$
(left column) and the patchy-particle reference (neighbour-list radius
$r_{\rm nbrs}=1.5$, right column), coloured by prescribed valency $f=1$--$6$.
(a,\,b)~Cost versus inverse temperature $\beta$: median with shaded interquartile band
per valency.  (c,\,d)~Cost versus local saturation $p$, one marker per simulated
condition.  For the VBC the cost is nearly flat in both $\beta$ and $p$ and rises only
weakly with valency, whereas the patchy cost is higher, far more scattered, and grows
steeply with valency and saturation.}
\end{figure*}

Figure~\ref{fig:wt_cost} shows this cost against inverse temperature $\beta$ and local saturation $p$ for the
standard $\mathrm{VBC}(1.5,2.0,0)$ and the patchy-particle reference. For the VBC the
wall time is nearly independent of both $\beta$ and $p$ and grows only weakly with
valency. The neighbour-count switch adds the same scalar work whatever the
thermodynamic state. The patchy-particle computational cost, by contrast, rises steeply with valency,
and exceeds the VBC cost already at low saturation, away from the
giant-component regime.

\begin{figure}[tb]
\centering
\includegraphics[width=\columnwidth]{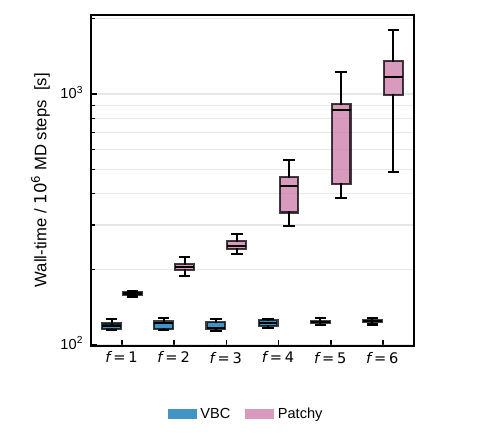}
\caption{\label{fig:wt_boxplot}
Neighbour-list benchmark on the tree/BVH backend ($\rho=0.05$, single GPU, buffer
$0.4$): wall time per $10^6$ MD steps for the standard $\mathrm{VBC}(1.5,2.0,0)$ (blue)
and the patchy reference ($r_{\rm nbrs}=1.5$, pink), shown as one box pair per
prescribed valency $f=1$--$6$.  The VBC cost is nearly valency-independent, whereas the
patchy cost grows steeply with valency.}
\end{figure}

Figure~\ref{fig:wt_boxplot} aggregates results over all simulated temperatures, showing
that the standard VBC stays efficient and nearly valency-independent, while the
patchy-particle reference approaches the VBC only at $f=1$ and slows down steadily as
the valency grows. Despite the larger neighbour-list radius, the VBC holds a uniform
advantage over the patchy-particle reference (Table~\ref{tab:performance}), reaching an almost
tenfold speed-up at the highest valency.

\begin{table*}[b]
\centering
\caption{\label{tab:performance}%
Median wall time per $10^6$ MD steps (seconds) on a single GPU, on the tree/BVH
neighbour-list backend, buffer $0.4$, for the standard VBC and the patchy reference
($r_{\rm nbrs}=1.5$) at each prescribed valency $f=1$--$6$, density $\rho=0.05$.
Medians are pooled over all simulated temperatures.}
\setlength{\tabcolsep}{6pt}%
\begin{tabular}{llcccccc}
\hline
Model & $\rho$ & $f=1$ & $f=2$ & $f=3$ & $f=4$ & $f=5$ & $f=6$ \\
\hline
$\mathrm{VBC}(1.5,2.0,0)$ & $0.05$ & $119$ & $122$ & $117$ & $122$ & $123$ & $124$ \\
Patchy                    & $0.05$ & $160$ & $204$ & $248$ & $429$ & $860$ & $1166$ \\
\hline
\end{tabular}
\end{table*}

\section{Applications of the valency-bounding correction}\label{sec:applications}

We now show two VBC applications in molecular dynamics that draw on very different
aspects of valency control.  The first lies in biophysics, where the correction lets
one simulate and equilibrate polymeric systems efficiently, reproducing behaviour that
otherwise demands far more expensive approaches.  The second uses valency control to remedy 
the penetrability of the soft repulsive core, thereby allowing soft-core potentials to be used across a wider range of binding strengths.  
Here the correction acts as a repulsion that grows adaptively whenever a
particle would violate its coordination bound.  In applications robust to the form of the repulsive core, 
it can therefore stand in for a hard or steeply repulsive core, keeping a large integration timestep and fast
equilibration.

\subsection{Modelling reentrant RNA aggregation}\label{sec:reentrant}

\begin{figure*}[t]
    \centering
    \includegraphics[width=\linewidth]{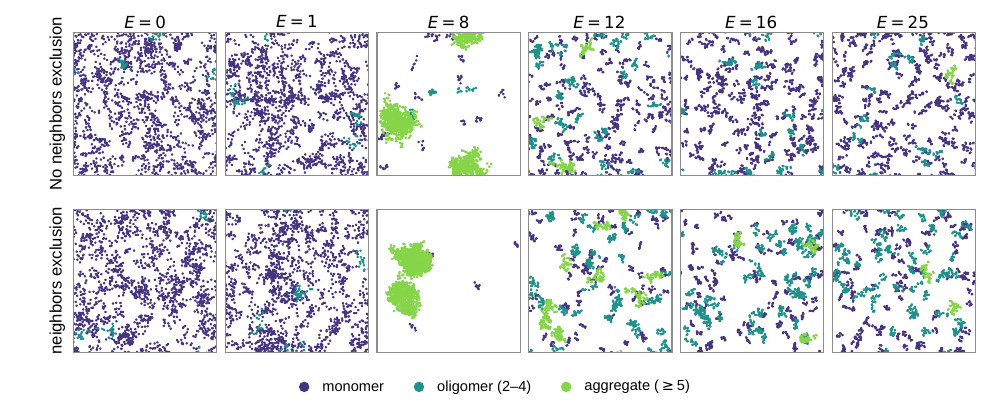}
    \caption{\label{fig:reentrant_snapshots}
    Simulation snapshots of the valency-one VBC sticker system across the reentrant
    transition, coloured by the cluster class of the chain (monomer, oligomer of
    $2$--$4$ chains, aggregate of $\ge5$ chains), for increasing sticker interaction
    strength $E_a$ (left to right).  Backbone-adjacent stickers are allowed to bind in
    the top row and excluded in the bottom row.  Aggregates form at intermediate $E_a$
    and redissolve into monomers and dimers at strong $E_a$.}
\end{figure*}

We choose reentrant RNA aggregation as a use case for the VBC.  Reentrant phenomena in
limited-valency systems have been characterised by analytical theory, simulation and
experiment~\cite{russo2011reentrant,roldan2013phase,bomboi2016re}. Limited valency 
is one of the key features accompanying reentrant transitions, so
reproducing them requires a model that controls the number of bonds per particle.  
This motivates our choice of RNA reentrant aggregation as a baseline
for showing that the VBC reproduces the expected behaviour at scalable computational
cost.

We focus on the transition described by Kimchi et al.~\cite{kimchi2023uncovering}.
Repeat-expanded RNA aggregates through self-complementary stickers, each of which pairs
with at most one partner. As predicted by an analytical multimer-enumeration
theory~\cite{kimchi2023uncovering}, RNA condensates form at intermediate sticker strength
but redissolve into monomers and dimers when the sticker attraction is made very strong.
Because each sticker binds a single partner, this regime can be enforced in simulation by
the VBC with $f=1$.

We model the strands as bead--spring chains whose stickers carry a prescribed valency
of one, separated by inert spacers, at a fixed strand concentration, and vary
the sticker interaction strength~$E_a$.  Following Kimchi et al.~\cite{kimchi2023uncovering}, we
classify every chain by the size $m$ of the cluster it belongs to: monomer ($m=1$),
oligomer ($2\le m\le4$) or aggregate ($m\ge5$).  Additionally, we compare scenarios in
which neighbouring stickers of the same strand are allowed or forbidden to bind.
Details of these simulations are given in Section~S3 of the ESI\,$\dag$.

\begin{figure}[t]
    \centering
    \includegraphics[width=0.95\columnwidth]{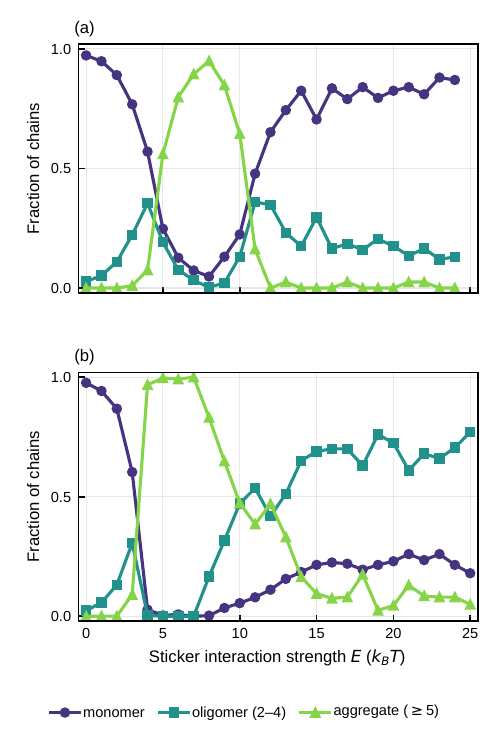}
    \caption{\label{fig:reentrant}
    Reentrant aggregation of valency-one VBC stickers, strand concentration $0.1$~mM.
    Fraction of chains that are monomers, oligomers ($2$--$4$ chains) or aggregates
    ($\ge5$ chains) versus sticker interaction strength~$E_a$, with sticker--sticker
    contacts adjacent along the backbone (a)~included or (b)~excluded.  The aggregate
    fraction peaks at intermediate $E_a$ and returns to zero at strong binding, 
    as the chains resolve into monomers and dimers.}
\end{figure}

The snapshots in Fig.~\ref{fig:reentrant_snapshots} show the transition directly:
aggregates dominate at intermediate $E_a$ and dissolve into monomers and dimers at strong
$E_a$.  Quantitatively, the aggregate fraction increases from zero to near unity at
intermediate $E_a$ and returns to low values at strong $E_a$, where the chains lock into
monomers and dimers that satisfy their bonds internally (Fig.~\ref{fig:reentrant}).
Whether monomers or oligomers survive at large binding energies is set by the exclusion
of binding between backbone-adjacent stickers, in agreement with the results of Kimchi
et al.~\cite{kimchi2023uncovering}.

Capturing the transition with the valency $f=1$ VBC is computationally feasible. Each
data point averages $2.5$ hours of single-GPU time for $7200$ particles over
$4\times10^{7}$ timesteps. This puts a full reentrant set of simulations, spanning the sticker
strength together with both backbone-binding scenarios, within a single-GPU budget of
about 150 GPU-hours, so the coarse-grained model turns a study that maps an entire
transition into a routine procedure.

\subsection{Adaptive repulsion for thermodynamic stability}\label{sec:stability}

Soft-core potentials are heavily used in large-scale coarse-grained simulations, because
they permit large integration time steps. This makes them a standard choice in
coarse-grained polymer~\cite{likos2001effective} and, specifically,
chromatin models~\cite{goloborodko2016chromosome,nuebler2018chromatin}.  However, a bounded
repulsive core in the presence of an attractive shell can cause thermodynamic
instability: the existence of the thermodynamic limit requires the total potential energy
$U=\sum_{i<j}V(r_{ij})$ of every configuration to satisfy the stability bound $U \ge -NB$
for some constant~$B$~\cite{fisher1966stability,ruelle1969statistical}.  Collapsing the
system into a cluster of radius comparable to $r_a$ makes an extensive fraction of all
$N(N-1)/2$ pairs attractive, while each overlapping pair pays only the finite penalty
$E_r$.  Beyond a threshold well depth the ground-state energy therefore scales as $-N^{2}$
and the free energy per particle diverges~\cite{fisher1966stability}.  In simulations this manifests 
as pathological accumulation of particles within an
extremely small volume.  Particles collapse into dense aggregates whose potential energy
decreases without bound, producing unphysical density singularities.

The conventional solution is to make the core impenetrable or steep, which restores
stability at the cost of the very timestep advantage that motivates soft cores.  The
VBC's switching function~$\Psi$ of Section~\ref{sec:potential} suppresses attraction once
the neighbour count exceeds the prescribed valency, creating an adaptive repulsion that
grows exactly where the collapse would develop, thus offering an alternative route to
thermodynamic stability.

For the VBC, a pair $(i,j)$ attracts only while $\Psi>0$, which requires $n_i < f+1$.
Every neighbour within the attractive range contributes at least $\omega_{\min}$ to the
count $n_i$, so an attracting particle carries fewer than $(f+1)/\omega_{\min}$
neighbours within $r_a$, and each attractive pair contributes at least $-E_a$, shared
between two particles.  The total attractive energy is then bounded:
\begin{equation}
\begin{aligned}
U_a &\;\ge\; -\,\frac{N\,(f+1)}{2\,\omega_{\min}}\,E_a \;\equiv\; -N B,\\[2pt]
\omega_{\min} &\equiv \min_{r \le r_a}\, \omega\bigl(r,\, r_\ell,\, r_h\bigr),
\end{aligned}
\label{eq:Hstable}
\end{equation}
which is the stability condition of Fisher and Ruelle~\cite{fisher1966stability}. The
bound holds provided $\omega_{\min}>0$. That is true as soon as the counting function assigns
nonzero weight to every particle within the attractive range. The choice
$r_h > r_a$ guarantees that at any well depth.  For the standard VBC the stronger choice
$r_\ell = r_a$ gives $\omega_{\min}=1$, and the bound reduces to $B=(f+1)E_a/2$.

If control over the coordination number is not required, the VBC can be used
to prevent collapse without interfering with binding. It is sufficient
to cap the number of overlapping cores, which is achieved by setting
$r_r \le r_h < r_a$. It leaves part of the shell attraction unaccounted
for and no longer controls the coordination number, but it still suppresses core
penetration. In particular, the choices $r_h = r_r$ and $r_h = \arg\min_r V_a(r)$
leave the short-range attraction unconstrained while creating an adaptive repulsion
that prevents soft-core penetration.

To test this picture, we simulate $N=1000$ particles at density $\rho=0.05$ with
$E_r=20$, $r_r=1$, $r_a=1.5$, and sweep the attractive depth $E_a$ from $0.1$ to $5$.
Four counter configurations are compared, ordered by decreasing coverage of the
attractive shell:
\begin{enumerate}[nosep,label=(\roman*)]
  \item Standard: $r_h=2.0$, $r_\ell=1.5$, $f=12$.  The counting shell fully encloses the attractive
    range ($\omega_{\min}=1$) and stability is guaranteed at every well depth.
  \item Partial: $r_h=1.25$, $r_\ell=0.75$, $f=0.5$.  Neighbours in the shell $1.25<r<r_a$ attract
    without being counted.
  \item Blind: $r_h=1.0$, $r_\ell=0.5$, $f=0.001$.  Only neighbours inside the repulsive core
    contribute to the count and the entire attractive shell is unmonitored.
  \item No VBC: $f=1000$.  The cap is never reached and the system reduces to the bare
    soft-core potential.
\end{enumerate}
In the partial and blind configurations the cap is made deliberately tiny, so that a single
counted neighbour already suppresses the attraction.  

\begin{figure}[!t]
    \centering
    \includegraphics[width=0.95\columnwidth]{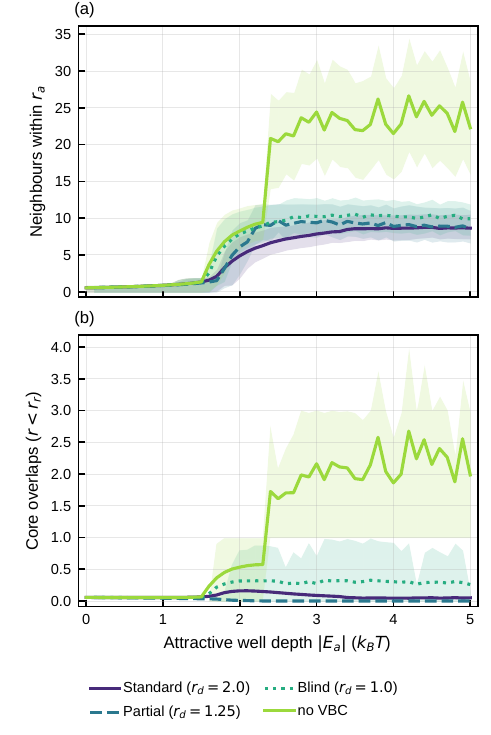}
    \caption{\label{fig:fr_stability}
    Signatures of Fisher--Ruelle instability in soft-core systems with varying counter
    coverage.  (a)~Mean contacts per particle within the attractive range~$r_a$.
    (b)~Mean penetrations per particle within the repulsive core~$r_r$.  Four counter
    configurations are compared, ordered by decreasing coverage of the attractive
    shell: the standard VBC with full coverage ($r_h=2.0$, dark purple solid), partial
    coverage ($r_h=1.25$, magenta dashed), blind coverage ($r_h=1.0$, pink dotted) and
    the bare soft-core potential without VBC (orange solid), with prescribed valencies
    $f=12$, $0.5$, $0.001$ and $1000$, respectively.  Shaded bands indicate
    the interquartile range across steady-state frames.  $N=1000$, $\rho=0.05$,
    $E_r=20$, $dt=0.01$.}
\end{figure}

Figure~\ref{fig:fr_stability} summarises the results.  Panel~(a) shows the mean number
of neighbours within the attractive range $r_a$ as a function of $E_a$.  The standard
configuration saturates below the prescribed valency.  The configurations with
$r_h<r_a$ accumulate progressively more contacts as $E_a$ grows, and the uncorrected
system reaches ${\sim}25$ neighbours per particle at $E_a=5$.  Panel~(b) shows the mean number of neighbours within the repulsive core ($r<r_r$).
Here the picture changes --- not only the standard configuration but also both
reduced-coverage configurations maintain near-zero core penetrations across the full
energy range, with the partial setup performing best.  Only the uncorrected system
shows extensive penetration, already at moderate attraction.

The VBC suppresses soft-core penetration for $r_r \le r_h < r_a$. Full coverage,
$r_h \ge r_a$, adds coordination control to the depth-independent bound of
Eq.~\eqref{eq:Hstable}, and every counting shell, including the exchange-tuned geometry
of Section~\ref{sec:lifetimes}, is thermodynamically safe for any short-range soft-core
potential.

\section{Conclusions}\label{sec:conclusions}

We have introduced the valency-bounding correction (VBC), a many-body modification of
short-range pairwise potentials that caps the coordination number of particles in
molecular dynamics simulations. The correction suppresses the attraction of a pair once
the smoothed neighbour count of either particle exceeds the prescribed valency. The
switching function is continuously differentiable and carries no angular terms, and the
forces evaluate in two passes over the neighbour list at the $O(kN)$ cost of a standard
pairwise potential. The VBC is available as an open-source GPU plugin for HOOMD-blue.

Across valencies $f=1$--$6$ the coordination number saturates at the prescribed value
on cooling. The cluster statistics depart from the Wertheim and Flory--Stockmayer
predictions, and the difference traces back to two properties of the correction. First,
all bonds of a particle share the same binding volume, so the saturation does not grow with
valency as it does for patchy particles. Second, the absence of angular constraints
permits ring formation, so bonds saturate within small cyclic clusters and the giant
component forms only at high saturation. These features could be used to simulate particles 
with flexible or surface-mobile
binding sites, while for systems bound by short-range directional bonds their effect on
the cluster statistics is compensated by a top-down adjustment of the attraction energy.

Systems with the VBC reach equilibrium quickly. An isotropic bond forms on contact, without an
orientational search, so the VBC fluid saturates its bonds close to an order of
magnitude faster than the patchy reference. The exchange-tuned
$\mathrm{VBC}(1.29,1.50,0.30)$ keeps bond lifetimes short at high saturation,
exchanging partners through the ordinary dynamics without dedicated swap moves. The
computational cost is nearly independent of valency, and at the highest valency it is
almost tenfold lower than that of the patchy-particle reference. The correction also
protects soft-core potentials from the Fisher--Ruelle collapse, so strong attraction can
be combined with large integration timesteps.

These properties make the VBC suited to problems that require large-system simulations or
extensive parameter scans: equilibration
of strongly bonded gels and networks, and simulation of multivalent biomolecular systems
such as sticker--spacer proteins, nucleic acids, and biomolecular condensates. The
reproduction of reentrant RNA aggregation at moderate computational cost illustrates this
capacity.

\section*{Author contributions}
V.D.: conceptualisation, methodology, investigation, formal analysis,
visualisation, writing. A.Gu.: software. A.Go.: conceptualisation,
supervision, funding acquisition, writing.

\section*{Conflicts of interest}
There are no conflicts to declare.

\section*{Data availability}
The valency-bounding correction is implemented as an open-source plugin for HOOMD-blue,
available at \url{https://github.com/glab-vbc/vbc_plugin}. All model definitions, simulation
parameters and analysis protocols required to reproduce the results of this work are
given in the main text and in the ESI\,$\dag$. Additional simulation outputs are
available from the corresponding author upon reasonable request.

\section*{Funding}
This work was supported by the Austrian Academy of Sciences, Austrian Science Fund (FWF) doc.fund program (B-112), and Austrian Science Fund (FWF) grant SFB F 8804-B "Meiosis".

\section*{Acknowledgements}
The computational results presented were obtained using the CLIP cluster of the Vienna
BioCenter. V.D.\ and A.Gu.\ are members of the Vienna BioCenter PhD Programme.

\bibliography{intro_references}

\end{document}